\pdfoutput=1

\documentclass[aps,pre,reprint,groupedaddress,longbibliography]{revtex4-1}

\usepackage{grffile}
\usepackage{cancel}
\usepackage{hyperref}
\usepackage{amssymb}
\usepackage{color}
\usepackage{amsmath}
\usepackage{bm}
\usepackage[english]{babel}
\usepackage{graphicx}
\usepackage{tikz}
\usetikzlibrary{positioning}
\usepackage{ifthen}
\usepackage{xstring}
\usepackage{afterpage}

\hypersetup{
  pdfproducer= none,  % producer of the document
  pdfcreator = none,  % producer of the document
  colorlinks = true,
  allcolors = black, %-- use this if you want to set all links to the same color
}

\newcommand{\Fig}[1]{Fig.~\ref{#1}}

\newcommand{\snake}{$\Sigma$}

\newcommand{\rr}{${\mathcal{R}}$}

\def\beq{\begin{equation}}
\def\eeq{\end{equation}}
\def\bea{\begin{eqnarray}}
\def\eea{\end{eqnarray}}

\def\beq{\begin{equation}}
\def\eeq{\end{equation}}
\def\bea{\begin{eqnarray}}
\def\eea{\end{eqnarray}}

\def\cal#1{\mathcal{#1}}
\def\eqq#1{Eq.~(\ref{#1})}
\def\eq#1{(\ref{#1})}

\def\f#1{Fig.~\ref{#1}}

\def\s#1{Section~\ref{#1}}

\def\c#1{~\cite{#1}}

\usepackage{array}
\newcolumntype{C}[1]{>{\centering\arraybackslash}p{#1}}

\begin{document}

%\title{Higher-order Ramachandran plots}
\title{The Ramachandran number: an order parameter for protein geometry}

\author{Ranjan V. Mannige}
\email[]{rvmannige@lbl.gov}
\affiliation{Molecular Foundry, Lawrence Berkeley National Laboratory, 1 Cyclotron Road, Berkeley, CA, U.S.A.}
\author{Joyjit Kundu}
\affiliation{Molecular Foundry, Lawrence Berkeley National Laboratory, 1 Cyclotron Road, Berkeley, CA, U.S.A.}
\author{Stephen Whitelam}
\email[]{swhitelam@lbl.gov}
\affiliation{Molecular Foundry, Lawrence Berkeley National Laboratory, 1 Cyclotron Road, Berkeley, CA, U.S.A.}

\date{\today}

\begin{abstract}
Three-dimensional protein structures usually contain regions of local order, called secondary structure, such as $\alpha$-helices and $\beta$-sheets. Secondary structure is characterized by the local rotational state of the protein backbone, quantified by two dihedral angles called $\phi$ and $\psi$. Particular types of secondary structure can generally be described by a single (diffuse) location on a two-dimensional plot drawn in the space of the angles $\phi$ and $\psi$, called a Ramachandran plot. By contrast, a recently-discovered nanomaterial made from peptoids, structural isomers of peptides, displays a secondary-structure motif corresponding to {\it two} regions on the Ramachandran plot [Mannige et al., {\em Nature} 526, 415 (2015)]. In order to describe such `higher-order' secondary structure in a compact way we introduce here a means of describing regions on the Ramachandran plot in terms of a single {\em Ramachandran number}, \rr, which is a structurally meaningful combination of $\phi$ and $\psi$. We show that the potential applications of \rr~are numerous: it can be used to describe the geometric content of protein structures, and can be used to draw diagrams that reveal, at a glance, the frequency of occurrence of regular secondary structures and disordered regions in large protein datasets. We propose that \rr~might be used as an order parameter for protein geometry for a wide range of applications. 
\end{abstract}

\maketitle

\usetikzlibrary{positioning}

\newcommand{\floor}[1]{\left \lfloor #1 \right \rfloor}
\newcommand{\round}[1]{\left \lfloor #1 \right \rceil }
\newcommand{\ceil}[1]{\left \lceil #1 \right \rceil }

\section{Introduction}
Many three-dimensional protein structures consist of regions of local order called secondary structure\c{Berg2006}. Consequently, the study of secondary structure has occupied a crucial role in structural biology\c{Chothia1997,Cooper2000,Berg2006,Linderstrom-Lang1952,Berg2006,Mannige2014b,Pauling1951a,Pauling1951,Linderstrom-Lang1952,Bragg1950,Eisenberg2003,Berg2006,Mannige2014b}. A key insight from this study is the recognition that the conformation of a protein backbone near a given amino acid residue can be specified largely by two dihedral angles, called $\phi$ and $\psi$, as shown in \f{fig1}(a) (a third angle, $\omega$, usually takes one of two values, defining {\em trans} and {\em cis} conformations\c{Pauling1951a,Pauling1951}). Ramachandran and co-workers deduced that peptide backbones inhabit only certain regions of dihedral angle ($\phi,\psi$) configuration space. Plots drawn in terms of this configuration space are called Ramachandran plots\c{Berg2006,Alberts2002,Mannige2014b,subramanian2001gn}, and they are among the most important innovations in structural biology, enabling immediate assessment of the geometric nature of protein structures\c{Laskowski1993,Hooft1997,Laskowski2003}. 

\begin{figure*}[t!]
 \includegraphics[width=0.9\textwidth]{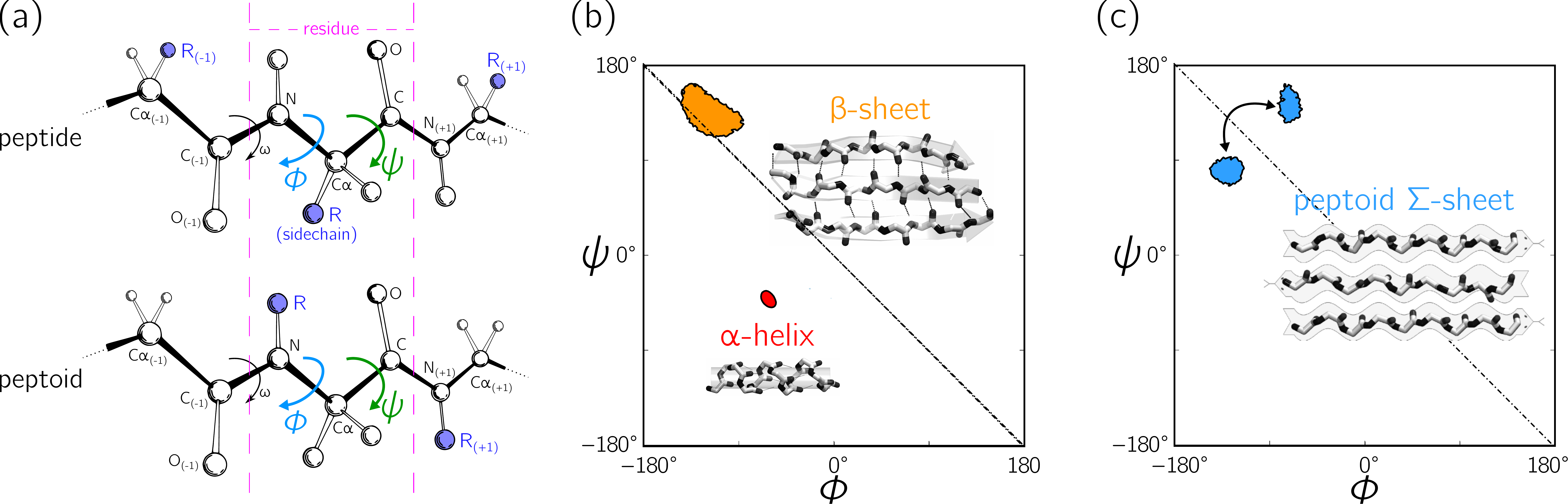}
  \caption{The Ramachandran plot is an important way of describing protein secondary structure. (a) The state of a residue within a peptide (top) and a peptoid (bottom) can be largely specified by the two dihedral angles $\phi$ and $\psi$. (b) Regular protein secondary structures, such as $\alpha$-helices and $\beta$-sheets, correspond to single diffuse regions on a plot drawn in terms of $\phi$ and $\psi$, called a Ramachandran plot (see Methods). (c) Peptoid $\Sigma$-sheets\c{mannige2015} harbor a secondary-structure motif in which backbone residues alternate between two regions on the Ramachandran plot. In order to describe each region in terms of a single number, so that the state of each residue in a backbone can be compactly indicated, we describe in this paper the development and properties of a structurally meaningful combination of $\phi$ and $\psi$ that we call the Ramachandran number, \rr. [Panel (a) was adapted from an image found on Wikimedia Commons (\href{https://commons.wikimedia.org/wiki/File\%3AProtein_backbone_PhiPsiOmega_drawing.jpg}{link}) by Dcrjsr (CC BY 3.0 (\href{http://creativecommons.org/licenses/by/3.0}{link})). 
  The contours in (b) and (c) represent regions within which 70\% of a secondary structure resides; see \s{scop}.]\label{fig1}}
  \includegraphics[width=0.9\textwidth]{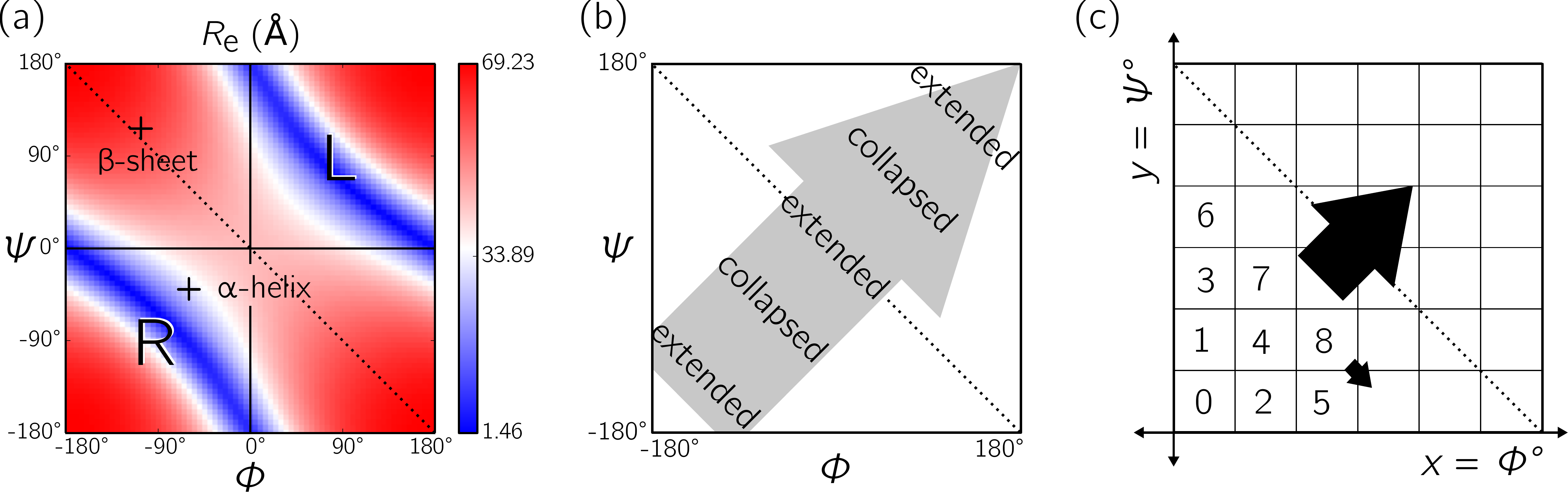}
  \caption{Physical trends within the Ramachandran plot suggest a way of describing regions of it with a single number. (a) First, the sense of residue twist changes from right-handed to left-handed as one moves from the bottom left of the Ramachandran plot to the top right. Second, contours (colored) of end-to-end polymer distance $R_{\rm e}$ (here calculated for a $20$-residue glycine) have a negative slope, resulting in the general trend shown in panel (b). Panel (c) indicates one method of indexing the Ramachandran plot so as to move from the region of right-handed twist to the region of left-handed twist with $R_{\rm e}$ changing as slowly as possible. This method provides the basis for the construction of the Ramachandran number, \rr.  \label{fig2}}
\end{figure*}

In general, residues that comprise particular protein secondary structures, such as $\alpha$-helices and $\beta$-sheets, correspond to distinct, localized regions on the Ramachandran plot; see \Fig{fig1}(b). However, the possibility of secondary structure built from more than one rotational state, i.e. more than one region on the Ramachandran plot, was introduced in 1951 by Pauling and Corey. They proposed a `pleated sheet' motif in which protein residues alternate between right- and left-handed forms of the $\alpha$-helix. While not yet seen in nature, simulations indicate that $\alpha$-pleated sheets can form as kinetic intermediates in unfolding processes\c{Armen2004,Daggett2006,Wu2010,babin2011alpha}. More generally, a broad range of protein structures could in principle be built from polypeptide motifs possessing two rotational states\c{Hayward2008,Hayward2014}. In the non-natural world, protein-mimetic polymers do form large-scale stable structures that simulation indicates harbor a secondary-structure motif built from more than one rotational state. The peptoid nanosheet\c{Nam2010} is a molecular bilayer that possesses macroscopic extent in two dimensions. It is made from peptoids, structural isomers of peptides. The nanosheet is flat because its constituent peptoid polymers are linear and untwisted, properties that result from the fact that backbone residues along each polymer alternate between two twist-opposed rotational states\c{mannige2015}. This secondary-structure motif, called a \snake-strand, corresponds to two regions on the Ramachandran plot, as shown in \f{fig1}(c).

To describe this structure and its possible generalizations it is convenient to be able to describe regions on the Ramachandran plot with a single number, so that the state of each residue along a polymer backbone can be compactly described. The desire for such a description is the motivation for this paper. We introduce in \s{number} a structurally meaningful combination of $\phi$ and $\psi$ that we call the Ramachandran number, \rr. Given a way of describing regions of the Ramachandran plot in terms of one number instead of two, one can then draw diagrams that give insight into protein geometry that is difficult to obtain by other means. In \s{properties} we show that \rr~can be used to assess in a compact manner the geometric content of protein structures, and can be used to draw diagrams that reveal at a glance the frequency of occurrence of regular secondary structures and disordered regions in large protein datasets. We also suggest that \rr~may be useful in the analysis of intrinsically-disordered proteins, whose three-dimensional structures are less well understood than those of globular proteins\c{Dunker2001,Dunker2008,Dunker2013,Mannige2014b}. We conclude in \s{conclusions}.

\section{One possible Ramachandran number}
\label{number}

One physical factor that suggests a compact way of describing the Ramachandran plot is the sense of residue twist implicit in the plot, which changes sign as one crosses the negative-sloping diagonal; see \f{fig2}(a). Structures in the bottom left-hand triangle of the Ramachandran plot have a right-handed (dextrorotatory) sense of twist, while structures in the top right-hand triangle have a left-handed (levorotatory) sense of twist\c{Berg2006,mannige2015}. This observation suggests an indexing system that proceeds from the bottom left of the plot to the top right of the plot. To gain insight into how this should be done, we built protein backbones with dihedral angles chosen from designated regions of the Ramachandran plot. We examined the behavior of the end-to-end distance $R_{\rm e}$ of polymers built in this way (the polymer radius of gyration behaves similarly). This behavior is shown in \f{fig2}(a,b).  The shapes of the contours of $R_{\rm e}$ suggest an indexing system that proceeds in a sweeping fashion across the Ramachandran plot, as shown in \f{fig2}(c), so that $R_{\rm e}$ changes as slowly as possible. Proceeding in this manner one moves from structures having one sense of twist to structures having the opposite sense of twist, with the degree of compactness of the backbone varying only in a gradual fashion. Thus the indexing system suggested graphically in \f{fig2}(c) is sensitive both to the twist state and the degree of compactness of the polymer backbone, allowing one to distinguish, for example, compact $\alpha$-helices from extended $\beta$-sheets, or nearly twist-free $\beta$ sheets from twisted loop regions.

\begin{figure}[t!]
\label{fig_pathology}
 \includegraphics[width=0.35\textwidth]{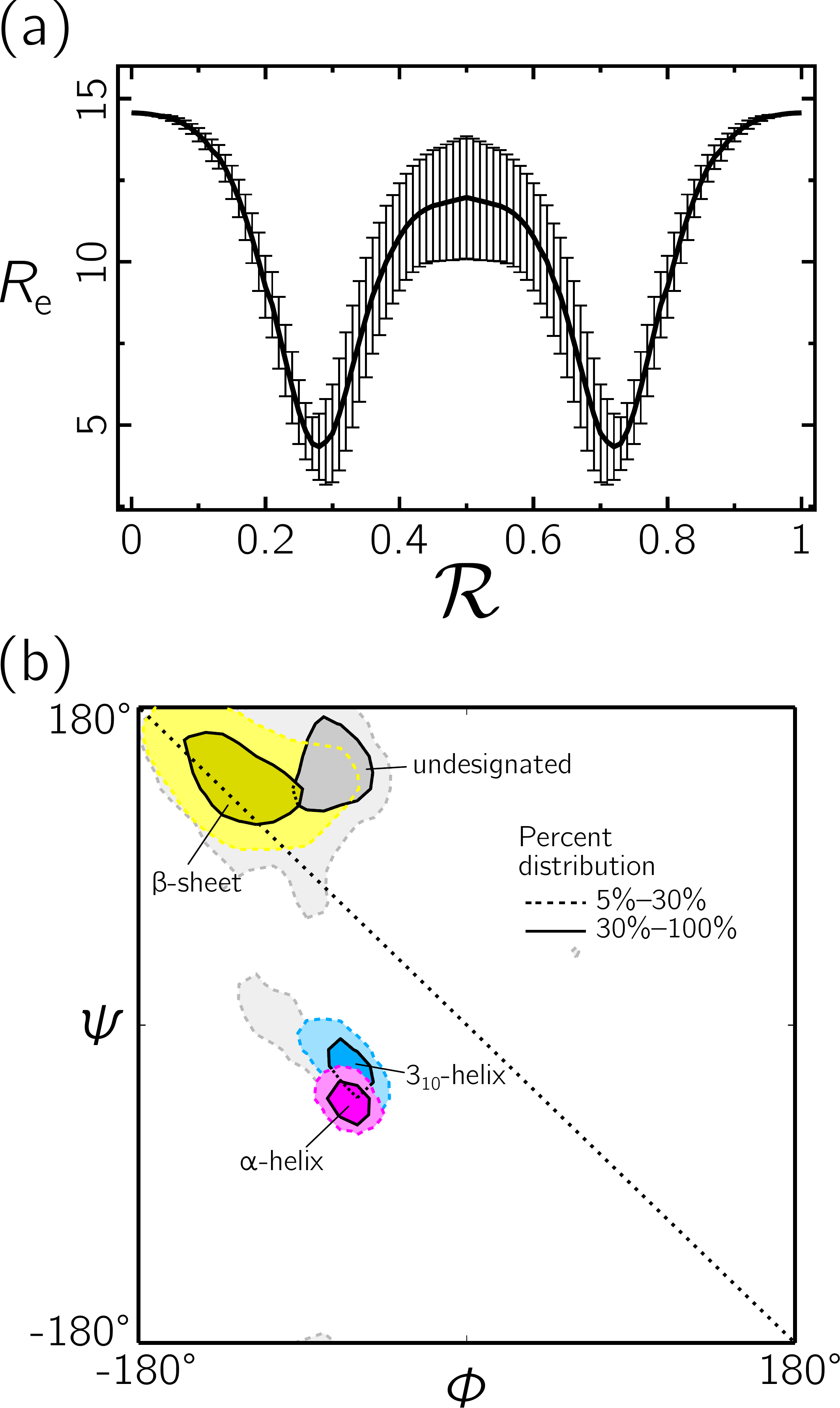}
  \caption{Potential pathologies of \rr~are avoided by the sparse occupancy of the Ramachandran plot. (a) We construct \rr~by slicing across the Ramachandran plot, which can cause points distant in dihedral angle space to be grouped together, the more so as we approach the negative-sloping diagonal (near \rr~$=0.5$). This grouping can be inferred by superposing the standard deviation (error bars) in polymer end-to-end distance on top of the mean value (smooth line) for hypothetical structures built from the relevant part of the Ramachandran diagram. (b) However, many structures distant in dihedral angle space but close in \rr~do not arise in proteins; the Ramachandran diagram is in general relatively sparsely occupied. Consequently, \rr~can resolve the major types of protein secondary structure, which can be inferred from the fact that lines parallel to the negative-sloping diagonal (marked), along which \rr~varies only slowly, can touch each region of known secondary structure (colored) individually. This sensitivity allows \rr~to function as an order parameter for protein geometry. [Data in (a) were calculated for a 5-residue peptoid; \rr~values are shown at discrete intervals of 0.01.] \label{r_pathogen}}
\end{figure}

To construct such an indexing system we take the Ramachandran plot axes to have the range $[-\lambda/2,\lambda/2]$ where $\lambda=360^\circ$\c{Berg2006,Alberts2002,Laskowski1993,Laskowski2003}. We divide the plot into a square grid of $(360^\circ \sigma)^2$ sites, where $\sigma$ is measured in reciprocal degrees. We shall show that it is straightforward to make $\sigma$ large enough that the error incurred upon converting angles from structures in the protein databank to Ramachandran numbers and back again is less than the characteristic error associated with the coordinates of structures in that database.

Given a choice of grid resolution $\sigma$ we define the real-valued normalized Ramachandran number
\beq
\label{ramachandran}
{\cal R}(\phi,\psi) \equiv  \frac{R(\phi,\psi)-R_{\rm min}}{R_{\rm max}-R_{\rm min}},
\eeq
where the unnormalized integer-valued Ramachandran number is
\beq
\label{ramachandran2}
R(\phi,\psi)  \equiv \round{\phi'} + \lambda' \round{\psi'}.
\eeq
In \eqq{ramachandran} we have defined $R_{\rm min} \equiv R(-180^\circ,-180^\circ)$ and $R_{\rm max} \equiv R(180^\circ,180^\circ)$. In \eqq{ramachandran2} the symbol $\round{x}$ means the integer closest to the real number $x$, and the parameter $\lambda' \equiv \round{\sqrt{2} \lambda\sigma}$. The coordinates $\phi' \equiv (\phi - \psi + \lambda)\sigma/\sqrt{2}$ and $\psi' \equiv (\phi+\psi + \lambda)\sigma/\sqrt{2}$ correspond to a clockwise rotation by 45$^\circ$, a shift, and a rescaling of the original coordinates $\phi$ and $\psi$; see \s{coords}.

The closest approximations to the original coordinates $\phi$ and $\psi$ that may be retrieved from $R$ are\c{Anderson1995}
\begin{align}
\tilde{\phi} &= \frac{1}{\sigma\sqrt{2}} \Big(\floor{R / \lambda'} + R\, \%\, \lambda' - \lambda' \Big), \label{backmap}\\
\tilde{\psi} &= \frac{1}{\sigma\sqrt{2}} \Big(\floor{R/\lambda'} - R\,  \%\,  \lambda' \Big),\label{backmap2}
\end{align}
where $\floor{x}$ is the largest integer smaller than the real number $x$, and $\alpha \, \% \,\beta$ is the remainder obtained upon dividing the integer $\alpha$ by the integer $\beta$. Equations \eq{ramachandran} to \eq{backmap2} define our mapping of the dihedral angles to the Ramachandran number, i.e. $(\phi,\psi) \to R \to$ \rr, and the subsequent approximate recovery of those angles, $R \to (\tilde{\phi},\tilde{\psi})$. We show in \s{sec_backmap} that this back-conversion can be done to within the characteristic precision of the protein databank.

By `slicing' across the Ramachandran plot we group together structures that might be relatively distant in dihedral angle space, the more so as we approach the negative-sloping diagonal (near \rr~$ = 0.5$). One consequence of this grouping is that the set of structures described by a small interval of \rr~displays a distribution of properties, such as end-to-end distance, as shown in \f{r_pathogen}(a). The mean of this distribution gives rise to a smoothly-varying trend, but the variance of this distribution is nonzero, and is largest near \rr~$ = 0.5$. Some unavoidable structural coarse-graining therefore occurs upon going from the Ramachrandran plot to the Ramachandran number. Despite this drawback, we shall show that \rr~can  function as an order parameter for protein geometry, in large part because the Ramachandran plot is in general relatively sparsely occupied: many hypothetical structures that possess distinctly different structural properties but that would be assigned similar Ramachandran numbers simply do not arise in the protein world. Consequently, \rr~can resolve the major classes of protein secondary structure, such as the $\alpha$ and $\beta$ motifs; see \f{r_pathogen}(b).

\afterpage{%
\onecolumngrid
\begin{figure*}
 \includegraphics[width=0.9\linewidth]{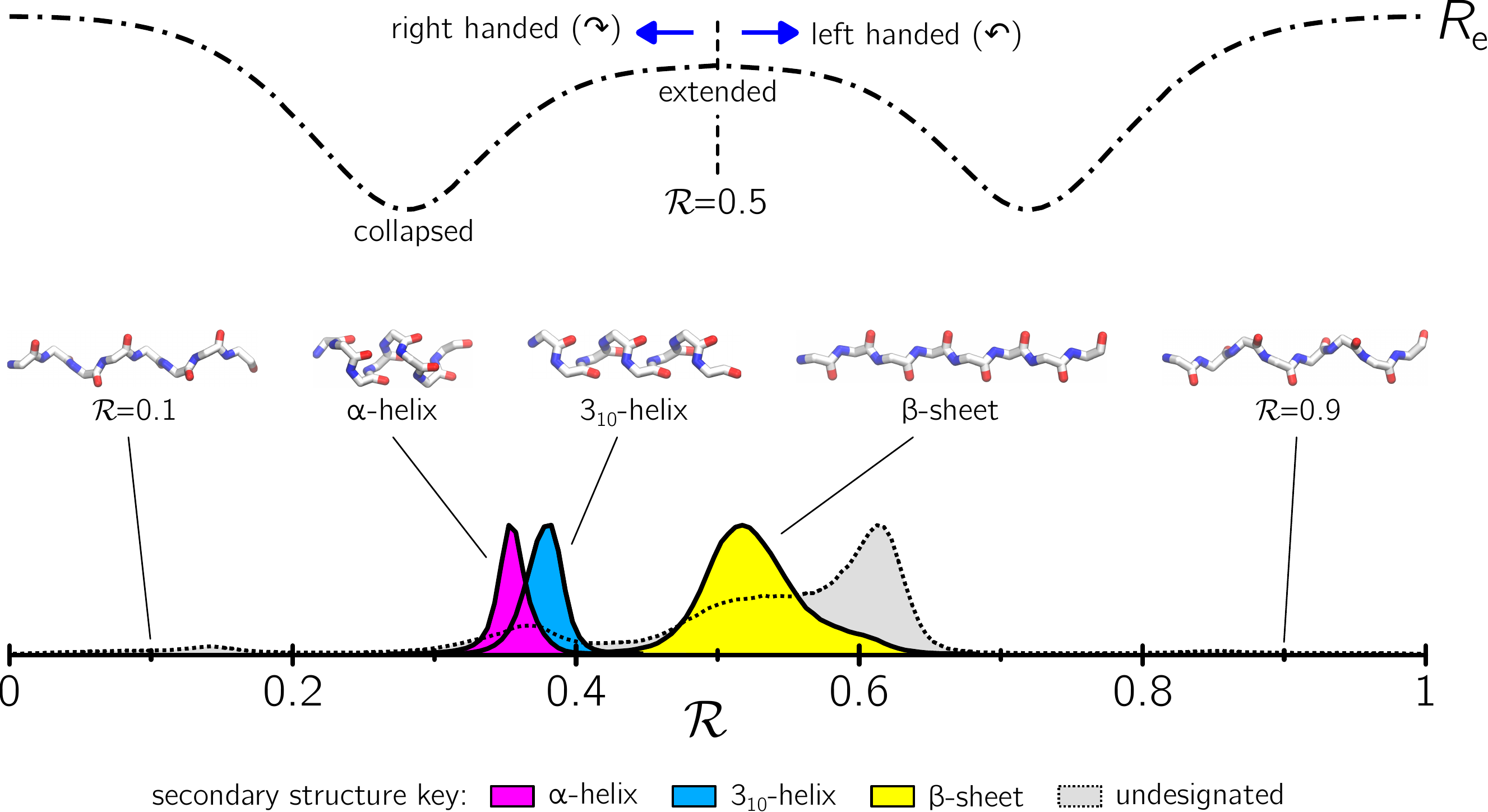}
  \caption{The indexing system defined by Equations \eq{ramachandran} and \eq{ramachandran2} collapses the Ramachandran plot into a single line, the Ramachandran number \rr. This number can act as an order parameter to distinguish secondary structures of different geometry, as shown (the overlap between distributions exists in the original Ramachandran plot representation; see \f{r_pathogen}(b)). Top: \rr~interpolates between regions of right-handed and left-handed twist, with polymer extension $R_{\rm e}$ varying smoothly throughout. \label{fig_line}}
 \includegraphics[width=0.9\linewidth]{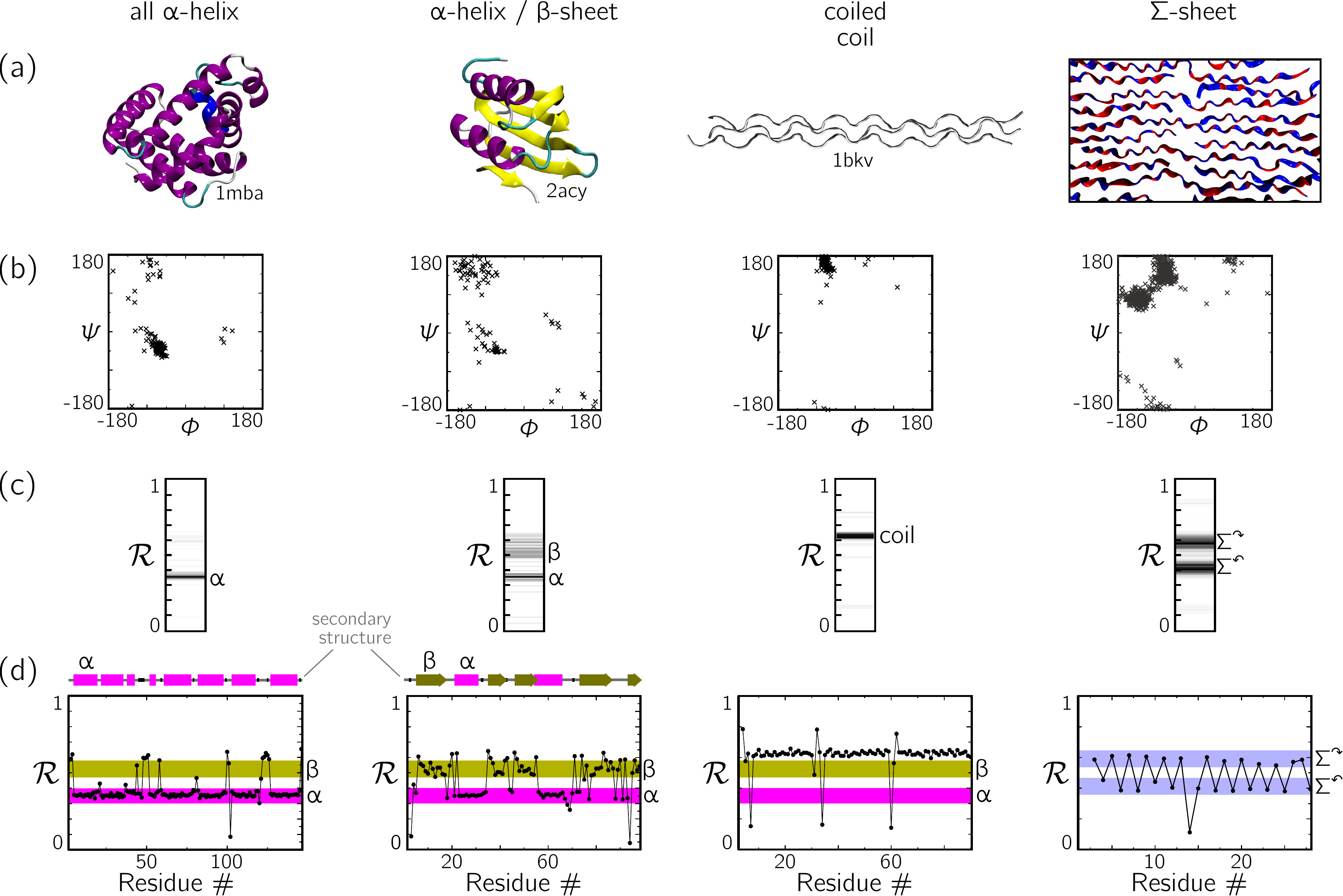}
  \caption{Four ways of looking at secondary structure: a) Molecular configurations; b) Ramachandran plot; c) Histogram (\rr-code) of Ramachandran numbers; and (d) \rr~as a function of residue number (for the \snake-sheet we have chosen a single polymer). Panel (c) provides a compact assay-by-geometry of the residues within molecular structures, while panel (d) shows that one can use \rr~to identify the spatial connectivity of domains of secondary structure within a polymer.\label{fig_collage}}
\end{figure*}
\clearpage
\twocolumngrid
}

\section{Properties and uses of \rr}
\label{properties}

The indexing system defined by Equations \eq{ramachandran} and \eq{ramachandran2} collapses the Ramachandran plot into a single line, the Ramachandran number \rr. As shown in \f{fig_line}, this number can act as an order parameter for types of polymer secondary structure. Given such an order parameter, we can then draw diagrams that reveal the abundance and spatial connectivity of different forms of secondary structure within polymers.

\begin{figure}[t!]
\includegraphics[width=\linewidth]{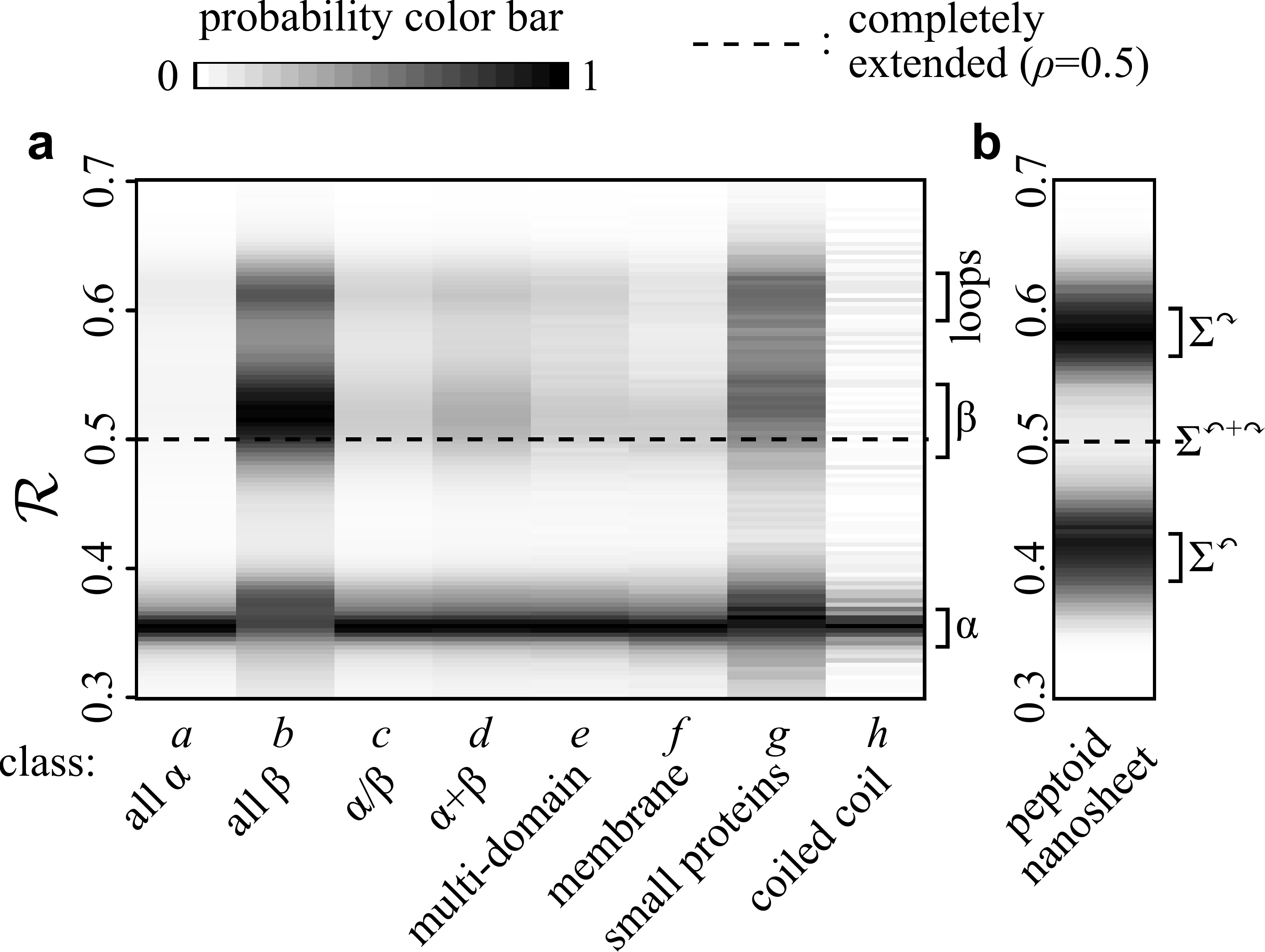}
\caption{(a) \rr-codes for the SCOP protein dataset reveal at a glance several geometric properties of the set. Each column represents a histogram of the indicated protein class, normalized so that the largest value is unity. A feature common to all classes is the prominence of $\alpha$-helices (\rr~$\approx 0.36$). Another common feature is the presence of loops that connect ordered secondary structure (\rr~$\approx 0.62$). Moreover, $\alpha$-helical regions are prominently visible in `all-$\beta$' proteins. (b) The \rr-code for a peptoid nanosheet shows two dominant rotational states, which coexist within a single secondary structure (see \f{fig_collage} and \f{fig_nanosheet}).
\label{fig_classes}}
\end{figure}

\begin{figure}[t!]
\includegraphics[width=\linewidth]{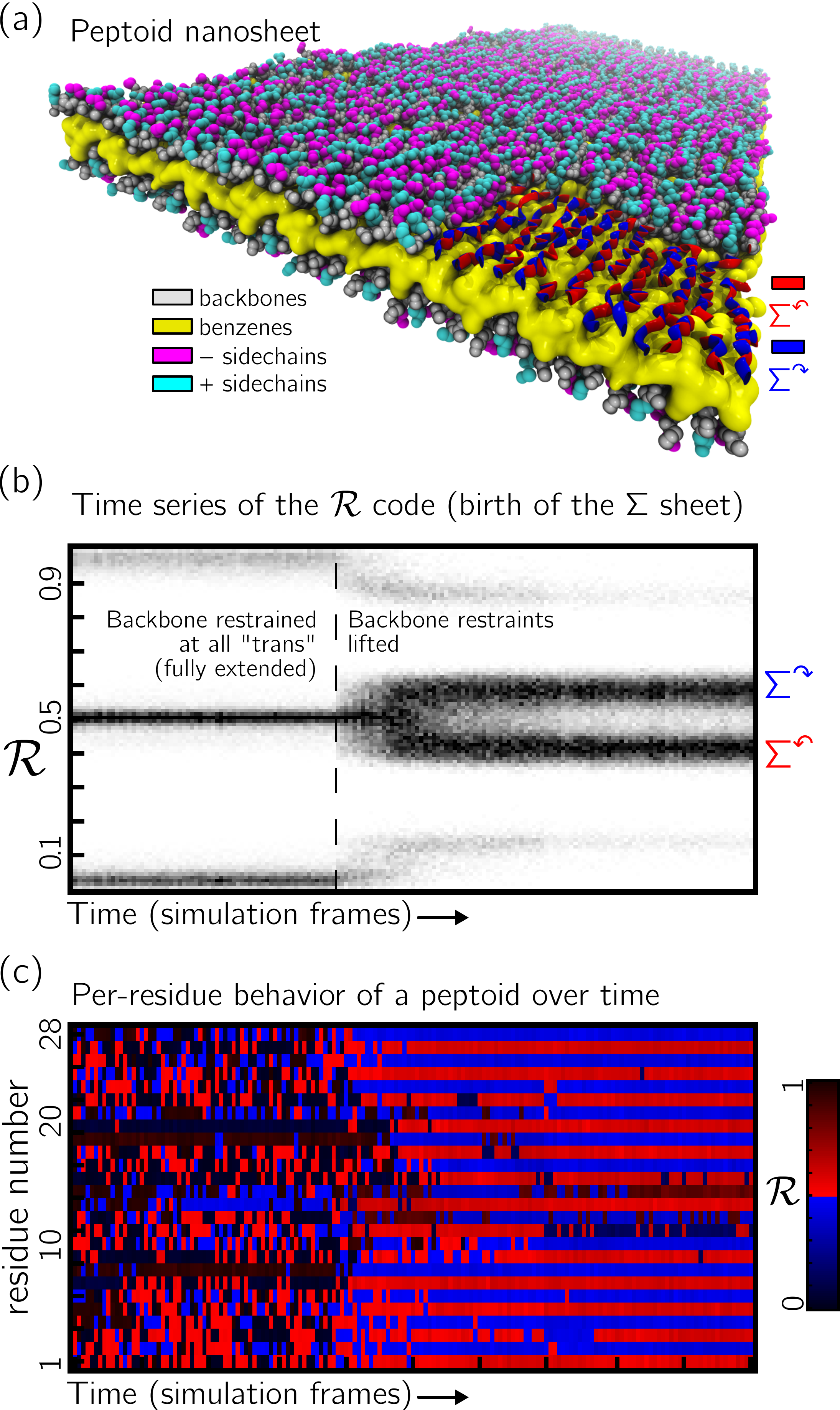}
 \caption{(a) Molecular dynamics simulations of the peptoid nanosheet\c{Nam2010,mannige2015} show the existence of the \snake-strand secondary structure motif, within which residues possess two distinct rotational states (colored red and blue in the bottom-right-hand cutaway). (b) A time series of the \rr-code of the bilayer shows the emergence (to the right of the vertical dotted line) of the \snake-strand motif within molecular dynamics simulations. Polymers in these simulations were initially fully extended, and adopted the \snake-strand motif upon relaxation of their backbone constraints. (c) Geometric state of each residue in one peptoid as a function of time, revealing the emergence of the \snake-strand structure and the subsequent fluctuations of individual residues on a nanosecond timescale. \label{fig_nanosheet}}
\end{figure}

In \f{fig_collage} we show four different molecular structures described in terms of (a) spatial configurations, (b) the Ramachandran plot, (c) a histogram (bar code or `\rr-code') of \rr-values, and (d) a plot of \rr~versus residue number (the structure of the coiled coil was deduced by a number of authors, Ramachandran among them\c{ramachandran1954structure,ramachandran1955structure,rich1955structure, bella1994crystal}). The \rr-code of panel (c) can be regarded as a way of assaying the residues of a protein by geometry, much as gel electrophoresis, which results in similar-looking pictures, is used to tell apart macromolecules by their size and charge. The plot of \rr~versus residue number in panel (d) reveals the spatial connectivity of distinct ordered domains. It shows the distinct segments of secondary structure ($\alpha$-helix and $\beta$-sheet) and loop regions in the proteins, and shows that the peptoid \snake-strand's residues alternate between twist-opposed rotational states. This representation makes clear that the two rotational states of the \snake-strand motif are incorporated within a single type of secondary structure; the Ramachandran plot alone does not distinguish between that outcome and the alternative, that the two rotational states exist within two distinct types of secondary structure.

\rr~can be used to compactly describe the abundance of secondary structure types with large protein datasets, as shown in \f{fig_classes}. There we show histograms of \rr~(`\rr-codes') for proteins belonging to distinct SCOP classes\c{Fox2014}. These diagrams identify a number of trends within this dataset. As expected, proteins belonging to classes `a' and `b' are rich in $\alpha$-helical (\rr~$\approx 0.36$) and $\beta$-sheet (\rr~$\approx 0.52$) regions, respectively. More surprisingly, $\alpha$-helical regions are abundant in all protein classes, even in the `all-$\beta$' class `b'. Loop regions (\rr~$\approx 0.62$) are also prominent; loops connect regions of ordered secondary structure. The \rr-code also highlights the symmetry of the peptoid backbone about the twist-free region \rr~$\approx 0.5$ (panel (b)).

\rr~can also be used in a time- and space-resolved way, as shown in \f{fig_nanosheet}. Here we show the results of molecular dynamics simulations of the peptoid nanosheet\c{Nam2010,mannige2015}, which reveal the existence of the \snake-strand secondary structure motif in which residues possess two distinct rotational states. A time series of the \rr-code of the bilayer (panel (b)) shows the emergence of the \snake-strand motif within molecular dynamics simulations via a breaking of the initially-imposed molecular symmetry. In panel (c), we show the geometric state of each residue in one peptoid as a function of time, revealing the emergence of the \snake-strand structure and the subsequent fluctuations of individual residues on a nanosecond timescale.

\section{Conclusions}
\label{conclusions}

The Ramachandran plot is central element of structural biology. We have introduced here a way of describing regions of the Ramachandran plot in terms of a single {\em Ramachandran number}, \rr, which is a structurally meaningful combination of $\phi$ and $\psi$. The are many possible ways of constructing such a number, and the one we have chosen is sensitive to the local twist state and degree of compactness of a polymer backbone. Given the ability to describe a two-dimensional space with a single number, one can draw diagrams that furnish insight into polymer structure that is difficult to obtain through other means. For instance, we have shown that \rr~can be used to describe the geometric content of protein and protein-inspired structures, in a space- and time-resolved way, and can be used to draw diagrams that reveal at a glance the frequency of occurrence of regular secondary structures and disordered regions in large protein datasets. We speculate that \rr~may also be useful in analyzing the behavior and evolution of intrinsically-disordered proteins (IDPs), important to e.g. the study of diseases\c{Babu2011}. Such proteins are less well-characterized than globular proteins\c{Berg2006}. IDPs spend substantial amounts of time in unfolded or disordered conformations\c{Dunker2001,Dunker2008,Dunker2013,Mannige2014b}, but may harbor local or transient regions of structure such as $\alpha$-helices\c{Geist2013}. The Ramachandran number described here may be a useful complement to existing bioinformatics metrics for IDP sequences\c{Dosztanyi2010} for understanding the behavior of these proteins in simulations\c{Staneva2012,Chong2013,Das2013,Mittal2014,Baruah2015} and experiments\c{Kosol2013,Sibille2012,Jensen2013,Konrat2014}. More generally, \rr~may be useful as an order parameter for polymer geometry for a wide range of applications.

\appendix

\renewcommand{\theequation}{A\arabic{equation}}
\renewcommand{\thefigure}{A\arabic{figure}}
\renewcommand{\thesection}{A\arabic{section}}

\setcounter{equation}{0}
\setcounter{section}{0}
\setcounter{figure}{0}

\section{Obtaining polymer (protein/peptide/peptoid) statistics.}
\label{scop}

The contours in \Fig{fig1}b and \Fig{r_pathogen}b describe the distribution of secondary structures in a Ramachandran plot, while \Fig{fig_line} represent histogram distributions of secondary structures on the Ramachandran line. 
To obtain statistics on secondary structures, a protein structure database was obtained from the Structural Classification of Proteins or SCOPe (Release 2.03)\c{Fox2014} 
that contains proteins with no more than 40\% sequence identity (downloaded from \verb+http://scop.berkeley.edu/downloads/pdbstyle/+\\
\verb+pdbstyle-sel-gs-bib-40-2.03.tgz+). Secondary structural elements such as $\alpha$-helices, $3_{10}$-helices and $\beta$-sheets were identified using the DSSP algorithm\c{Zhao2005,Kabsch1983,Joosten2011}.

\Fig{fig_classes}a represents \rr-codes for entire classes of proteins. We utilized the SCOP classification system and the individual proteins from each class were extracted from the SCOP dataset described above. Altogether, there were 8560 proteins that were amenable to analysis from this database.

The distribution for the $\Sigma$-sheet on a Ramachandran Plot (\Fig{fig1}c) and in an \rr-code (\Fig{fig_collage}, \Fig{fig_classes}b) were obtained from a 50 nanosecond interval of a molecular dynamics trajectory\cite{mannige2015}. \Fig{fig_nanosheet}(b,c) describes a trajectory of the same system before and after the symmetry-enforcing backbone restraints were lifted. This process was part of a molecular dynamics equilibration step used in Ref.\c{mannige2015}.

The end-to-end distances for glycine peptides of 20 residues (\Fig{fig2}(a)) and 5 residues (\Fig{r_pathogen}(a)) were generated using the PeptideBuilder library\c{Matthew2013} and analyzed using BioPython\c{Cock2009}.

\section{Coordinate transformation used to obtain \rr}
\label{coords}
\begin{figure}
\label{fig_coord}[t]
 \includegraphics[width=0.48\textwidth]{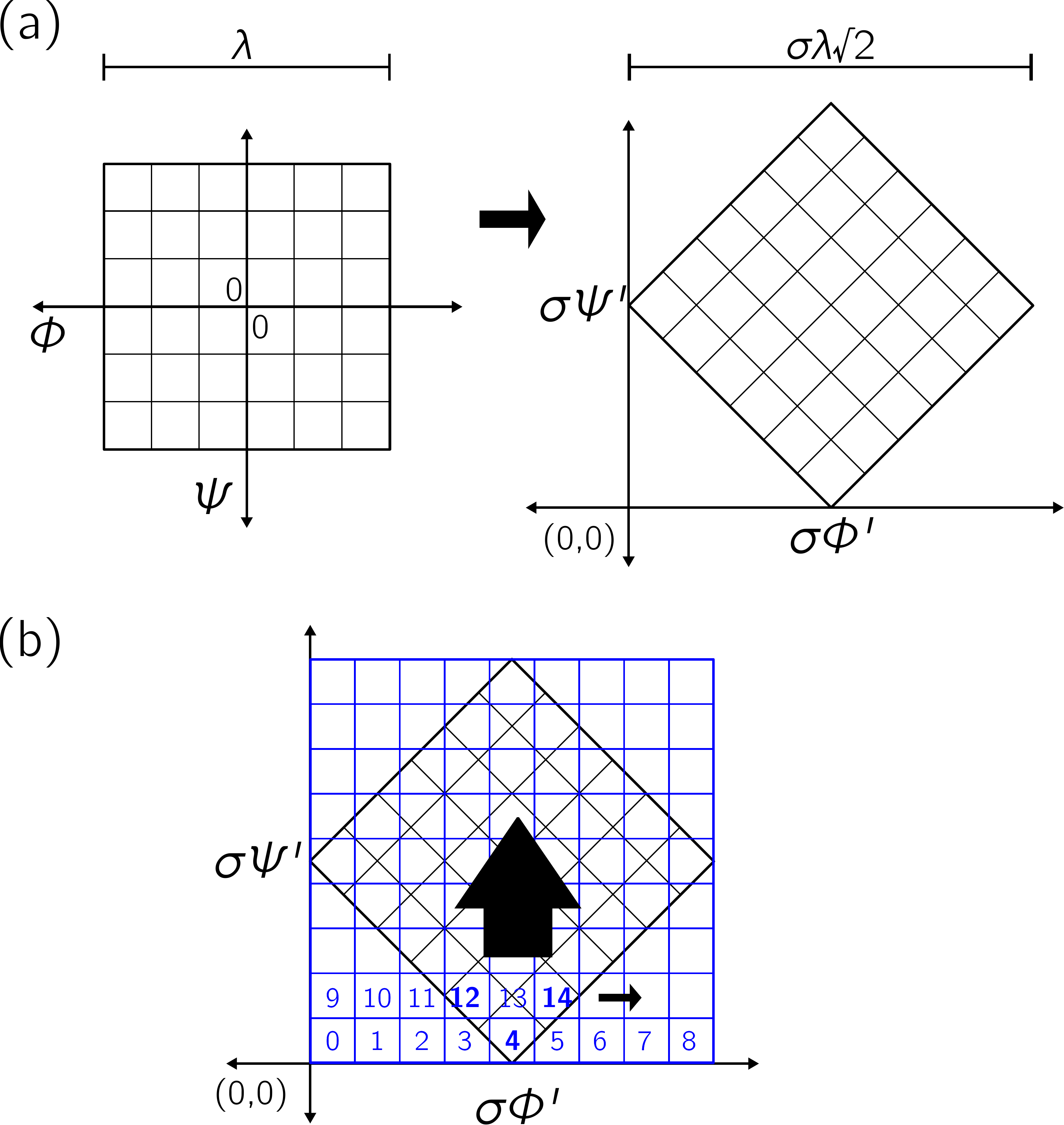}
  \caption{(a) Coordinate transformation applied to the Ramachandran plot in order to compute the Ramachandran number \rr. A rotation, shift, and rescaling of $\phi$ and $\psi$ results in a representation in which horizontal cuts run roughly along contours of polymer extension; see \f{fig2}. (b) The indexing system defined by \eqq{ramachandran2}, where bold numbers are those that fall on the original Ramachandran plot.}
\end{figure}

Equations \eq{ramachandran}--\eq{backmap2} were obtained by rotating the Ramachandran plot so that contours of constant polymer extension $R_{\rm e}$ (see \f{fig2}) lie roughly horizontal in the new coordinate representation. To this end we define
\beq
\left(\begin{matrix} \phi'\\ \psi' \\ \end{matrix} \right) \equiv
\frac{\sigma}{\sqrt{2}} \left(\begin{matrix}   1 & -1 \\ 1 & 1 \\ \end{matrix} \right)    
\left(\begin{matrix} \phi\\ \psi \\ \end{matrix} \right)+
 \frac{\lambda \sigma}{\sqrt{2}}\left(\begin{matrix} 1\\ 1 \\ \end{matrix} \right),
\eeq
so that $\phi'$ and $\psi'$ are obtained by rotating the original coordinate system $(\phi,\psi)$ clockwise by 45$^\circ$, shifting the resulting coordinates linearly (so that the new coordinates are non-negative), and rescaling the result by the grid resolution $\sigma$. \f{fig_coord}(a) shows graphically this transformation. Indexing the new coordinate system according to \eqq{ramachandran2} corresponds to the counting system shown in \f{fig_coord}(b). To undo the transformation \eq{ramachandran2} approximately we compute
\bea
\tilde{\phi'} &=& R \, \% \,  \lambda',  \label{back1}\\
\tilde{\psi'} &=& \floor{R /\lambda'} \label{back2}.
\eea
We then insert \eq{back1} and \eq{back2} into the equations $\tilde{\phi'} \equiv \left(\tilde{\phi} - \tilde{\psi} + \lambda\right)\sigma/\sqrt{2}$ and $\tilde{\psi'} \equiv \left(\tilde{\phi}+\tilde{\psi} + \lambda\right)\sigma/\sqrt{2}$, and solve these for the closest approximations $\tilde{\phi}$ and $\tilde{\psi}$ to the original angles $\phi$ and $\psi$. The results are Equations \eq{backmap} and \eq{backmap2}.

\section{Recovering dihedral angle values from the Ramachandran number} 
\label{sec_backmap}
It is convenient to be able to retrieve from the Ramachandran number a good approximation to the dihedral angles used to calculate it. This can be done for a range of choices of grid resolution $\sigma$. We took 8560 protein structures obtained from SCOP\c{Fox2014}; see \s{scop}. For a given protein we took the dihedral angles associated with each residue, and used these to compute the 3D protein structure (given values of the $\omega$ dihedral angle). We carried out the conversion from dihedral angles to Ramachandran number, defined by \eqq{ramachandran2}, and from this used Equations \eq{backmap} and \eq{backmap2} to obtain an approximation to the original dihedral angles. We used these approximate angles (with the original values of the $\omega$ dihedral angle) to compute the 3D structure of the protein. We then calculated the root-mean-squared deviation (RMSD) between the original and recovered sets of angles and heavy-atom positions, shown in \f{fig_backmap}. For a range of values of grid resolution $\sigma$ we find these RMSD values to lie well within the 1\AA~characteristic of the protein databank. For the calculations done in this paper we took $\sigma = 10^5$ reciprocal degrees.

\begin{figure}[b!]
  \includegraphics[width=0.7\linewidth]{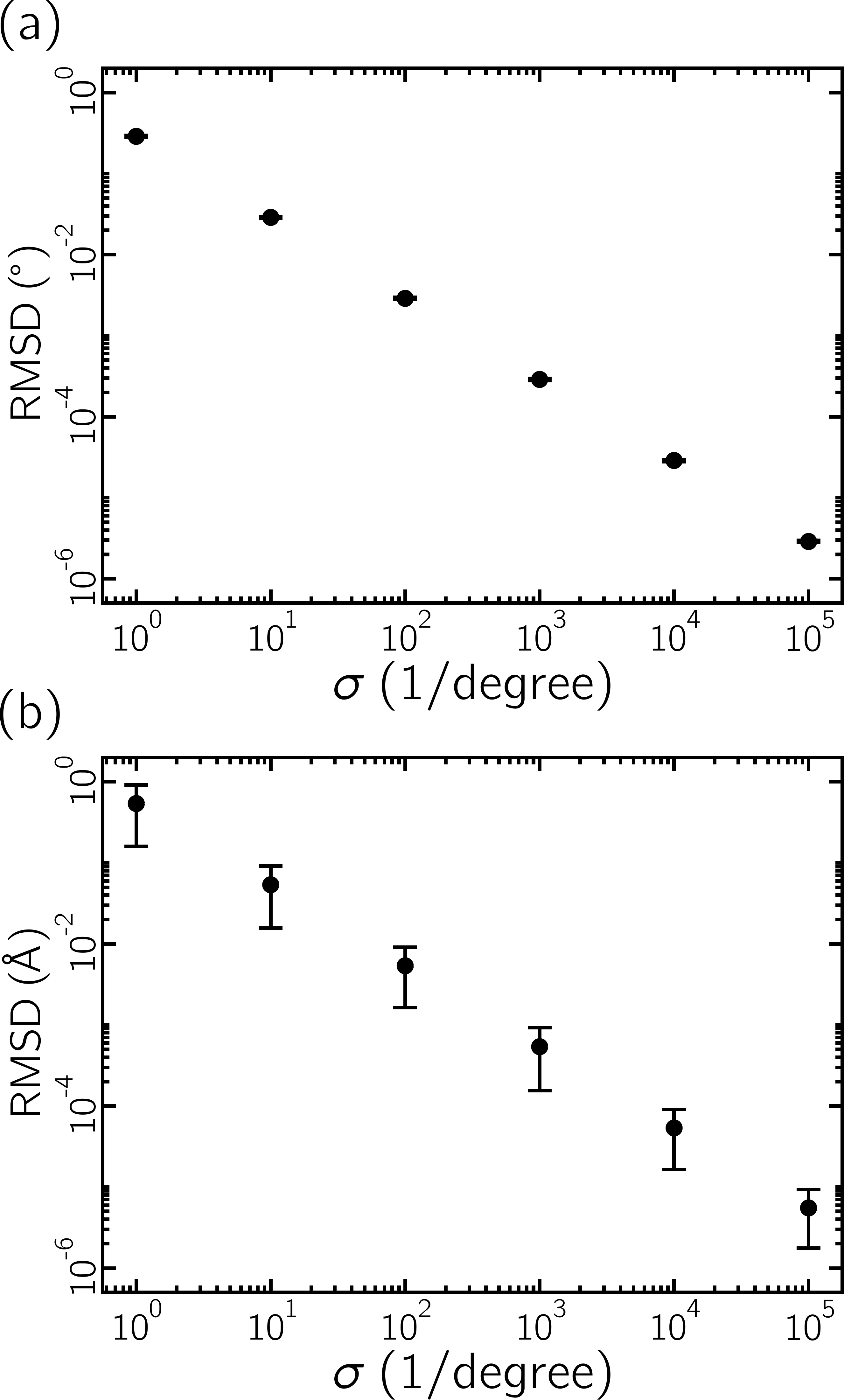}
  \caption{Dihdral angles converted to Ramachandran numbers can be recovered only approximately, but the error incurred during this back-mapping can be made much smaller than the standard error (typically 1\AA) associated with structures in the protein databank. Here we show the root-mean-squared-deviation (RMSD) in dihedral angles (a) and in protein $\alpha$-carbon spatial coordinates (b) generated upon taking 8560 protein structures obtained from SCOP\c{Fox2014}, converting their dihedral angles to Ramachandran numbers, and recovering approximately those dihedral angles using \eqq{backmap} and \eqq{backmap2}. The parameter $\sigma$ indicates the grid resolution used to calculate $R$; see \eqq{ramachandran2}.\label{fig_backmap}}
\end{figure}

\begin{acknowledgments}
RVM and SW were supported by the Defense Threat Reduction Agency under contract no. IACRO-B0845281. This work was done at the Molecular Foundry at Lawrence Berkeley National Laboratory (LBNL), supported by the Office of Science, Office of Basic Energy Sciences, of the U.S. Department of Energy under Contract No. DE-AC02-05CH11231.\end{acknowledgments}

%\bibliography{all.bib}


\onecolumngrid
% Resetting the figure numbers
\setcounter{figure}{0}
% Resetting the page numbers
\setcounter{page}{1}
% Making the figure numbers begin with "S"
\makeatletter 
\renewcommand{\thefigure}{S\@arabic\c@figure}
\renewcommand{\thesection}{S\@arabic\c@section}
\renewcommand{\theequation}{S\@arabic\c@equation}
\renewcommand{\thetable}{S\@arabic\c@table}
% Making the page numbers begin with "S"
\renewcommand{\thepage}{S\@arabic\c@page}
\makeatother

%\section{Supplementary Information}

\setcounter{secnumdepth}{3}
\renewcommand{\thetable}{S\arabic{table}}
\renewcommand{\thefigure}{S\arabic{figure}}
\renewcommand{\thesection}{S\arabic{section}}
\renewcommand{\thesubsection}{\thesection.\arabic{subsection}}
\renewcommand{\thesubsubsection}{\thesubsection.\arabic{subsubsection}}
\setcounter{table}{0}
\setcounter{figure}{0}
\setcounter{section}{0}
\setcounter{subsection}{0}
\setcounter{subsubsection}{0}

\renewcommand{\thesection}{S}
%\section{Appendix/Supplimentary Information}

\setcounter{section}{0}
\setcounter{subsection}{0}
\setcounter{subsubsection}{0}
\renewcommand{\thesection}{S\arabic{section}}

\end{document}